\documentclass[prl,twocolumn,showpacs,preprintnumbers,amsmath,amssymb,floatfix]{revtex4}

\usepackage[]{graphicx}             
\usepackage{bm}                     
\usepackage{tabularx}

\newcommand{\scP}{\mathcal{P}}

\begin{document}

%
%
\title{Lineshape distortion in a nonlinear auto-oscillator near generation threshold: Application to spin-torque nano-oscillators}

\author{Joo-Von Kim}
\email{joo-von.kim@ief.u-psud.fr}
\author{Q. Mistral}
\author{C. Chappert}
\affiliation{Institut d'Electronique Fondamentale, UMR CNRS 8622, Universit{\'e} Paris-Sud, 91405 Orsay, France}
\author{V. S. Tiberkevich}
\author{A. N. Slavin}
\affiliation{Department of Physics, Oakland University, Rochester, MI 48309, USA}
\date{\today}                                           

%
%
\begin{abstract}
The lineshape in an auto-oscillator with a large nonlinear frequency shift in the presence of thermal noise is calculated. Near the generation threshold, this lineshape becomes strongly non-Lorentzian, broadened, and asymmetric. A Lorentzian lineshape is recovered far below and far above threshold, which suggests that lineshape distortions provide a signature of the generation threshold. The theory developed adequately describes the observed behavior of a strongly nonlinear spin-torque nano-oscillator.
\end{abstract}


\maketitle

%
%

Thermal noise plays a key role in the dynamics of auto-oscillatory systems, especially for the nano-sized systems in which the characteristic energy of auto-oscillation is comparable with the thermal noise energy $k_BT$. One important consequence of noise is the broadening of the power spectrum. It is well-established, for example, that white Gaussian noise leads to a broadened Lorentzian shape of the power spectrum both far below and far above the generation threshold in conventional auto-oscillators (for which the frequency is weakly dependent on the amplitude)~\cite{Hempstead:PR:1967}. In addition, the presence of noise also blurs the threshold between the damped and auto-oscillatory states. 

In the immediate vicinity of the auto-oscillation threshold, a non-Lorentzian spectrum appears: the lineshape can be described by a sum of several Lorentzian profiles corresponding to several simultaneously excited \emph{statistical} modes. These modes have the same central frequency for a conventional auto-oscillator and, therefore, the contributions of higher-order modes do not lead to large qualitative changes in the overall spectral line profile~\cite{Hempstead:PR:1967}. In contrast, the central frequencies of the partial Lorentzian profiles are shifted in frequency by different amounts in a \textit{nonlinear} auto-oscillator in which the frequency is strongly dependent on the amplitude. In the vicinity of the auto-oscillation threshold at which these shifts are of the order of the oscillation linewidth, the frequency nonlinearity leads to important qualitative changes in the lineshape: the spectral line is asymmetric and significantly broadened. 

An example of an auto-oscillating system with strong frequency nonlinearity is the spin-torque nano-oscillator (STNO), which is of considerable interest at present. This oscillator consists of two (``free" and ``fixed") ferromagnetic layers separated by a nonmagnetic (metallic or dielectric) spacer. When driven by a spin-polarized direct current, an effective negative damping results in the ``free'' layer due to the spin-transfer effect~\cite{Slonczewski:JMMM:1996, Berger:PRB:1996}. For a sufficiently large direct current, the induced negative damping overcomes the natural positive damping of the ``free" layer, and auto-oscillations of magnetization of this layer result~\cite{Tsoi:PRL:1998, Kiselev:Nature:2003, Rippard:PRL:2004}. These and related phenomena are subjects of intensive research at present due to their possible applications in microwave nano-electronics.

In this Letter, we present a theory of lineshape distortion in auto-oscillators with a strong nonlinear frequency shift in the presence of thermal noise. We show that line broadening and line asymmetry are key signatures of the threshold region, and we argue that such features provide a simple way to determine the threshold in experiment. This point is especially significant for STNOs in which the influence of noise on the device performance is of a qualitative importance. The theory is compared to recent experiments on STNO nanopillars for which good qualitative agreement is found.

Our theory is based on a classical Hamiltonian formalism for spin-waves in magnetic multilayers with the spin-transfer effect~\cite{Slavin:ITM:2005,Rezende:PRL:2005}. It is assumed that only \emph{one} spin-wave mode -- the mode having the lowest relaxation rate -- is excited at the auto-oscillation threshold, and dominates the oscillation dynamics both below and just above this threshold. Using this assumption and following Refs.~\onlinecite{Slavin:ITM:2005, Kim:PRB:2006} we can reduce the Landau-Lifshitz equation with Gilbert damping, which describes the magnetization dynamics of the STNO ``free" layer, to a stochastic nonlinear oscillator equation for the complex amplitude $c$ of the single excited spin-wave mode ($|c| \leq 1$)~\cite{Tiberkevich:APL:2007},
\begin{equation}
\frac{d c}{d t} +i\omega(p)c +\Gamma_{\rm+}(p)c -\Gamma_{\rm
-}(I, p)c = f(t). \label{eq:EOM}
\end{equation}
Here, $p=|c|^2$ in the dimensionless oscillation power which determines the polar angle $\theta$ of the magnetization precession in the ``free" layer $\theta \equiv \arccos(M_z/M_0) = \arccos(1 - 2|c|^2)$, where $M_0$ is the length of the magnetization vector and $M_z$ is the projection of this vector on the direction of stationary equilibrium magnetization $\bf z$ (see Ref.~\onlinecite{Slavin:ITM:2005} for details). The oscillation frequency $\omega(p)$ is given by $\omega(p) \simeq \omega_0 + N p$, where $\omega_0$ is the (linear) mode frequency, and the coefficient $N$ describes a nonlinear frequency shift. Typically for STNOs, $|N| \gtrsim \omega_0$, which is the key feature distinguishing STNO from other conventional ``quasi-linear" auto-oscillators for which $N\simeq0$. The second term in Eq.~(\ref{eq:EOM}), $\Gamma_+(p) \simeq \Gamma_0(1 + Q p)$, represents natural \emph{positive} spin-wave damping, where $\Gamma_0$ is the linear relaxation rate and $Q$ is the phenomenological dimensionless parameter characterizing nonlinearity of the damping~\cite{Tiberkevich:PRB:2007}. The third term, $\Gamma_{\rm -}(I, p) = \sigma I(1 - p)$, describes current-induced nonlinear  damping (which is \emph{negative} for small oscillation amplitudes), where $I$ is the bias current, and $\sigma$ is the spin-polarization efficiency defined by  Eq.~(5) in Ref.~\onlinecite{Slavin:ITM:2005}.

The function $f(t)$ in (\ref{eq:EOM}) is a stochastic term that describes the influence of the thermal fluctuations. We take $f(t)$ to represent a white Gaussian process with zero mean and spectral properties of the form $\langle f(t)f^*(t')\rangle = 2\Gamma_0\eta \delta(t-t')$. The strength of the stochastic term is given by the product of the linear relaxation rate $\Gamma_0$ in the ``free" layer and the thermal noise energy $\eta = k_B T / (\lambda\omega_0)$, where $\lambda$ is the constant that relates dimensionless oscillation power $p=|c|^2$ to the energy of the system $E(p) = \lambda \omega_0 p$. In our case $\lambda = VM_0/\gamma$, where $V$ is the volume of magnetic material involved in auto-oscillations, $M_0$ is the saturation magnetization, and $\gamma$ is the gyromagnetic ratio. This choice of noise amplitude $\eta$ ensures that the system, in the absence of spin-transfer, relaxes towards the equilibrium thermal noise level $\langle E(c) \rangle = k_B T$ at the rate $\Gamma_0$~\cite{Kim:PRB:2006, Tiberkevich:APL:2007}.

It should be stressed that the nonlinear oscillator equation~(\ref{eq:EOM}) is rather general, and applies not only to spin-transfer-induced dynamics in magnetic multilayers, but also to the dynamics of any nonlinear auto-oscillating system operating in a single-mode regime.

In experiments involving STNOs the magnetization oscillations driven by spin-transfer torque are characterized by the spectral density of the associated voltage oscillations. It has been shown elsewhere that the autocorrelation function of experimentally measured voltage oscillations $v(t)$ is directly proportional to the autocorrelation function of the corresponding spin-wave oscillations  to lowest order~\cite{Kim:PRB:2006}, i.e. $\langle v(t) v(0) \rangle \propto {\rm Re}\! \left[ \langle c(t) c^{*}(0) \rangle \right]$. In what follows, we demonstrate how to calculate the correlation function $K(t) \equiv \langle c(t) c^{*}(0) \rangle$ or the power spectrum,
\begin{equation}\label{def-S}
 S(\omega) \equiv \int_{-\infty}^\infty K(t) e^{i\omega t} dt,
\end{equation}
of spin-wave oscillations resulting from the stochastic dynamics described by Eq.~(\ref{eq:EOM}) at any magnitude of the driving bias current $I$.

It is possible to obtain a linearized form of Eq.~(\ref{eq:EOM}) far above and far below threshold, from which the spin-wave correlation functions can be evaluated exactly. In the subcritical regime, $\zeta \equiv I/I_{\rm th} \ll 1$ (where $I_{\rm th} \equiv \Gamma_0/\sigma$ is the threshold current at which the auto-oscillations start), all nonlinearities can be ignored and Eq.~(\ref{eq:EOM}) leads to a Lorentzian power spectrum $S(\omega)$ with linewidth
\begin{equation}
\Delta \omega = \Gamma_0 - \sigma I.
\label{eq:linewidth_linear}
\end{equation}
This linear dependence has recently been observed in spin-valve nanopillars~\cite{Krivorotov:Science:2005} and tunnel junctions~\cite{Petit:PRL:2007} in the low-current regime. Far above the threshold ($\zeta > 1$), the lineshape is also Lorentzian with a linewidth governed by phase noise~\cite{Kim:PRB:2006,Kim:2007},
\begin{equation}
\Delta \omega = \Gamma_0 \, ( k_B T/E_0 ) \left[ 1 + (  N/\Gamma_{\rm eff})^2  \right],
\label{eq:linewidth_nonlinear}
\end{equation}
where $E_0 = \lambda\omega_0 p =\lambda\omega_0 (\zeta-1)/(\zeta + Q)$ is the average energy of the stable auto-oscillation in STNO, $N = d\omega(p)/dp$  is the nonlinear frequency shift coefficient, and $\Gamma_{\rm eff} = \Gamma_+(p)/dp - d\Gamma_-(p)/dp$ is the effective nonlinear damping (see Ref.~\onlinecite{Kim:2007} for details).

Near threshold, the coupling between amplitude and phase fluctuations leads to a non-Lorentzian power spectrum. To demonstrate this, we use the polar coordinates $c(t) = r(t) \exp[i \phi(t)-i\omega_0 t]$ and linearize (\ref{eq:EOM}) about the steady-state trajectory, $r_0 = \sqrt{(\zeta-1)/(\zeta+Q)}$, just above threshold, $r(t) = r_0 + a(t)$~\cite{Kim:2007},
\begin{eqnarray}
\dot{a}(t) + 2 \Gamma_{\rm eff}r_0^2 a(t) &=& {\rm Re}[\tilde{f}(t)], \\
\dot{\phi}(t) &=& \frac{1}{r_0} {\rm Im}[\tilde{f}(t)]- 2N r_0 a(t),
\end{eqnarray}
where $\tilde{f}(t) = f(t)e^{i\omega_0 t-i\phi(t)}$ represents a white Gaussian noise with the same statistical properties as $f(t)$. While the amplitude fluctuations $a(t)$ are subjected to a white-noise forcing only, $a(t)$ itself becomes a colored-noise source for the phase fluctuations $\phi(t)$, since $\langle a(t)a(t') \rangle \sim \exp(-2\Gamma_{\rm eff}r_0^2|t-t'|)$. The stochastic process of the phase variable $\phi(t)$ is therefore driven by a colored-noise source $a(t)$ and a white noise source $\tilde{f}(t)$, which clearly leads to a non-Lorentzian power spectrum if we take the STNO to be a simple phase oscillator.

This linearization scheme is not accurate in the threshold region because the steady state amplitude $r_0$ is comparable to the magnitude of the amplitude fluctuations. However, instead of solving the full {\em nonlinear} stochastic problem (\ref{eq:EOM}), it is more tractable to solve the corresponding {\em linear} Fokker-Planck equation, which describes the time evolution of the probability density function (PDF) $\scP(t,c;c_0)$. A similar approach has been used to study the problems of thermally-assisted magnetization reversal in the presence of spin-transfer torque~\cite{Li:PRB:2004} and telegraph noise in bi-stable current-induced precessional states~\cite{Gmitra:PRL:2007}. The PDF $\scP(t,c;c_0)$ gives the probability of the auto-oscillator to have the complex amplitude $c$ at the time $t$, given the amplitude at the zero moment of time ($t=0$) was $c_0$. From this non-stationary PDF $\scP(t,c;c_0)$, the correlation function $K(t)$ for $t>0$ can be obtained as
\begin{equation}\label{def-K}
    K(t) = \int d^2c \int d^2c_0 \scP(t, c; c_0)\scP_0(c_0) c\, c_0^*
,\end{equation}
where $\scP_0(c)$ is the stationary PDF, i.e. $\scP_0(c) = \lim_{t\to\infty}P(t,c;c_0)$. For $t<0$ one can use relation $K(t) = K^*(-t)$.

The linear Fokker-Planck equation can be obtained from (\ref{eq:EOM}) in a standard way~\cite{Risken:1989}, which in polar coordinates is given by
\begin{multline}
\label{eq:FP}
    \frac{\partial\scP}{\partial t} = %
        \frac{1}{r}\frac{\partial}{\partial r}\left\{\left[\Gamma_+(r^2) - \Gamma_-(I,r^2)\right]r^2\scP\right\}%
        +N r^2 \frac{\partial\scP}{\partial\phi} \\
        + \eta\Gamma_0\left[ \frac{1}{r}\frac{\partial}{\partial r}\left(r\frac{\partial\scP}{\partial r}\right) + \frac{1}{r^2}\frac{\partial^2\scP}{\partial\phi^2} \right] \equiv \hat{L}\scP.
\end{multline}
The stationary solution $\scP_0$ of Eq.~(\ref{eq:FP}) does not depend on the time $t$ and angle $\phi$, i.e. $\scP_0 = \scP_0(r)$, and can be found explicitly in a general case~\cite{Tiberkevich:APL:2007}. Eq.~(\ref{eq:FP}) is a linear equation that does not depend on time $t$ and phase $\phi$ explicitly. Therefore, its solution $\scP(t,c;c_0) = \scP(t,r,\phi;r_0,\phi_0)$ can be written as a
series
\begin{equation}\label{expansion}
    \scP(t,r,\phi;r_0,\phi_0) = \sum_{n,\mu}a_{n,\mu}(r_0,\phi_0)P_{n,\mu}(r)e^{-\Lambda_{n,\mu}t-in\phi}
,\end{equation}
where $P_{n,\mu}(r)$ are normalized eigensolutions of the problem
    $\hat{L}_nP_{n,\mu} = -\Lambda_{n,\mu}P_{n,\mu}$,
with
    $\hat{L}_n \equiv \hat{L}    (\partial_r,-in)$.
The eigenvalues $\Lambda_{n,\mu}$ and coefficients $a_{n,\mu}(r_0,\phi_0)$ satisfy the initial conditions for the PDF: $\scP(0,r,\phi;r_0,\phi_0) = (1/r)\delta(r-r_0)\delta(\phi-\phi_0)$. In the expansion (\ref{expansion}), the index $n$ describes the $\phi$-dependence $\sim \exp(-in\phi)$ of the particular statistical mode $P_{n,\mu}$, while the index $\mu$ enumerates different modes having the same $\phi$-dependence.

With the normalized solutions $Q_{n,\mu}(r)$ of the adjoint eigenproblem,
   $ \hat{L}_n^+Q_{n,\mu} = -\Lambda_{n,\mu}^{*} Q_{n,\mu}$,
and using Eqs.~(~\ref{def-K}) and (\ref{expansion}) with Eq.~(\ref{def-S}),
a series expansion for the power spectrum $S(\omega)$ can be obtained in the form,
\begin{equation}
S(\omega) = \sum_\mu{\frac{F_{1,\mu} \, {\rm Re}(\Lambda_{1,\mu})}{[\omega-{\rm Im}(\Lambda_{1,\mu})]^2+ [{\rm Re}(\Lambda_{1,\mu})]^2}},
\label{eq:lineshape}
\end{equation}
whereby each contribution is weighted by the factor
 $   F_{1,\mu} = \int_0^\infty dr \int_0^\infty dr' \, (r r')  P_{1,\mu}(r) \scP_0(r) Q_{1,\mu}^{*}(r').$
Each term in (\ref{eq:lineshape}) is associated with one statistical eigenmode $(n,\mu)$. Only the $n=1$ modes give contributions to the second-order correlator $K(t)$ and power spectrum $S(\omega)$. Statistical modes with other phase indices $n$ would be necessary to calculate the higher-order correlators, such as $\langle[c^*(t)c(0)]^2\rangle$. The partial lineshape associated
with each statistical mode is Lorentzian, with the central frequency and linewidth given by the imaginary and real parts of the eigenvalues $\Lambda_{1,\mu}$, respectively. The result in (\ref{eq:lineshape}), therefore, represents an expansion of the non-Lorentzian spectrum in terms of partial Lorentzians, whereby each mode $\mu > 1$ represents an additional relaxation process (and time scale) due to the additional colored-noise source. 

\begin{figure}
\includegraphics[width=8.5cm]{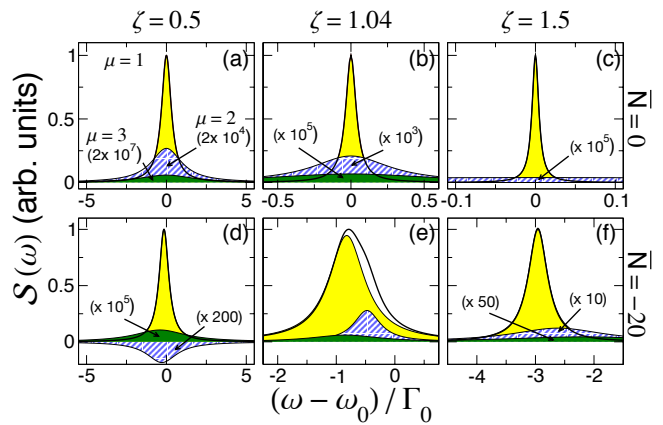}
\caption{\label{fig:spectral_lines}(Color online) Power spectrum, $\mathcal{S}(\omega)$, for $\eta = 10^{-3}$ and $Q = 2$. The dominant $\mu$ modes contributing to the overall spectral line(thick black curve) are shown for two values of frequency nonlinearity $\bar{N}$ at different supercriticality $\zeta = I/I_{\rm th}$.}
\end{figure}

A comparison between the lineshape distortion of a nonlinear ($\bar{N} = N/\Gamma_0= -20$) and linear ($\bar{N} = 0$) oscillators is presented in Fig.~\ref{fig:spectral_lines}.
The noise level considered is $\eta = 10^{-3}$, which corresponds approximately to $T = $ 300 K for the nanopillar system in \cite{Mistral:APL:2006} and $T = $ 400 K for the nanocontact system in \cite{Rippard:PRL:2004} with an in-plane applied field of 0.6 T. Far below and above threshold, the spectral lines are largely Lorentzian and are well described by a single $\mu=1$ mode for which the weighting factor $F_{1,1}$ is at least two orders of magnitude larger than the others, as one can clearly see in Fig.~\ref{fig:spectral_lines}. A significant deformation of the spectral line occurs near and just above threshold, as discussed earlier, which appears as large contributions from the higher-order eigenmodes in (\ref{eq:lineshape}). For the nonlinear oscillator, there is an additional lineshape asymmetry due to the different central frequencies of the partial Lorentzians. As each Lorentzian represents a stochastic process related to a different amplitude process (resulting in a different relaxation rate), there is an associated frequency shift due to the nonlinearity that is different for each process. A detailed description of these processes will be presented elsewhere.

\begin{figure}
\includegraphics[width=8.5cm]{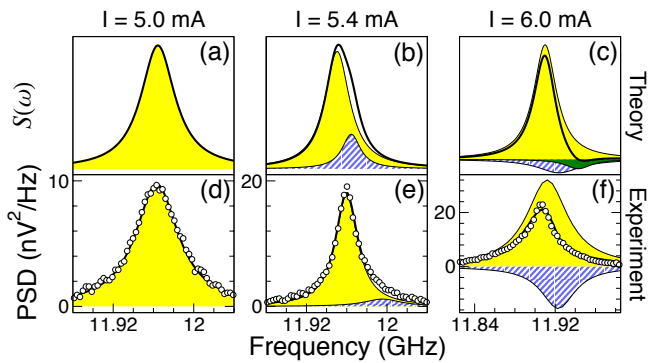}
\caption{\label{fig:comp_lineshape}(Color online) Comparison of theoretical and experimental spectral lines near threshold ($I_{\rm th} =$ 5.2 mA). Thick solid lines in (a)-(c) represent spectral lines as predicted by Eq.~(\ref{eq:lineshape}), with the first dominant $\mu$ modes shown as shaded curves. Circles in (d)-(f) are experimental data at $T = $ 225 K taken from Ref.~\onlinecite{Mistral:APL:2006}, with thick solid lines representing a double Lorentzian fit. Shaded curves are individual Lorentzian profiles used in the fits.}
\end{figure}

Salient features of such lineshape asymmetry have been observed in a recent experiment~\cite{Mistral:APL:2006}. In Fig.~\ref{fig:comp_lineshape}, the experimental spectral lines are fitted and compared with the theoretical lineshapes predicted by the Fokker-Planck theory.
Based on the linewidth and power variations of the experimental data we estimated the experimental critical current to be $I_{\rm th} = $ 5.2 mA, which allows us to compare the theoretical and experimental lineshapes at the same supercriticality $\zeta$. Below threshold ($I =$ 5.0 mA, $\zeta \approx$ 0.96) the experimental data are well described by a single Lorentzian curve (Fig.~\ref{fig:comp_lineshape}e), which is consistent with the theoretical curve (Fig.~\ref{fig:comp_lineshape}a). Just above threshold ($I =$ 5.4 mA, $\zeta \approx$ 1.04), a slight asymmetry appears in the spectral line whereby a second Lorentzian peak is required for a reasonable fit. The presence of a second mode is also seen in the theoretical curve, with a good qualitative agreement in the relative positions and amplitudes of the two modes (Fig.~\ref{fig:comp_lineshape}b). At a moderate above-threshold current ($I =$ 6.0 mA, $\zeta \approx$ 1.15), the lineshape distortion becomes more pronounced (Fig.~\ref{fig:comp_lineshape}f), and this asymmetry is qualitatively reproduced by the theory (Fig.~\ref{fig:comp_lineshape}c).

\begin{figure}
\includegraphics[width=7cm]{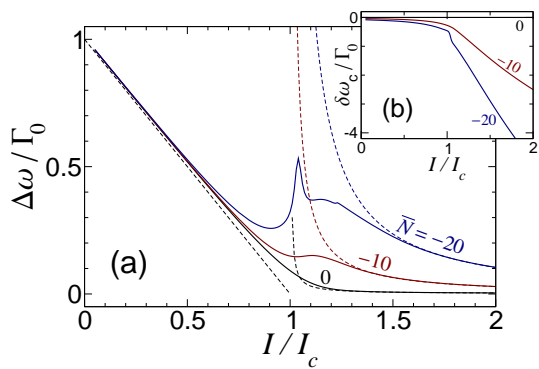}
\caption{\label{fig:fitted_spectra}(Color online) (a) Scaled linewidth and (b) frequency shift, $\delta \omega_c = \omega_c - \omega_0$, as a function of supercriticality from Lorentizian fits to spectra computed for different $\bar{N}=N/\Gamma_0$ for $Q= 2$, $\eta_0 = 10^{-3}$. Dashed lines in (a) correspond to Eqs.~(\ref{eq:linewidth_linear}) and (\ref{eq:linewidth_nonlinear}).}
\end{figure}

Another way of quantifying the lineshape distortion is provided by single Lorentizian fits to the spectra which yield the central frequency $\omega_c$  and the linewidth $\Delta\omega$. The results of fits of the peaks computed from Eq.~(\ref{eq:lineshape}) for different values of $\bar{N}$ are shown in Fig.~\ref{fig:fitted_spectra}.
The overall linewidth $\Delta\omega$, in general, decreases with the increase of current (Fig.~\ref{fig:fitted_spectra}a), except in the region near the generation threshold at which a local maximum of the linewidth can appear due to the lineshape asymmetry. Depending on the magnitude of the nonlinear frequency shift $N$, this maximum can be several times larger than the generation linewidth in both subcritical and supercritical regimes. Good agreement between the numerically and analytically calculated linewidths is found far below and far above the threshold in all cases, as demonstrated by the dashed lines representing Eqs.~(\ref{eq:linewidth_linear}) and (\ref{eq:linewidth_nonlinear}), respectively. Experimentally measured variations of the STNO linewidth with supercriticality are in good qualitative agreement with the developed theory (see Ref.~\onlinecite{Tiberkevich:APL:2007}). The variation of the oscillation frequency (Fig.~\ref{fig:fitted_spectra}b) is relatively slow below the threshold and exhibits a fast quasi-linear decrease with current above the threshold due to the nonlinear frequency shift (determined by the coefficient $N$). Only the frequency redshifts ($N < 0$) have been presented here for the sake of brevity; analogous behavior is predicted for the frequency blueshifts.

In summary, we have presented a theory of lineshape distortion for auto-oscillators with strong frequency nonlinearities. The Lorentzian lineshape of the power spectrum becomes asymmetric in the threshold region. Good qualitative agreement with recent experimental data on the spin-transfer nano-oscillators is demonstrated.

\begin{acknowledgments}
This work was in part supported by the MURI grant W911NF-04-1-0247 from the US Army Research Office, the contract W56HZV-07-P-L612 from the U.S. Army TARDEC, RDECOM, the grant ECCS-0653901 from the National Science Foundation of the USA, the Oakland University Foundation, and the European Communities program IST under Contract No. IST-016939 TUNAMOS.
\end{acknowledgments}

\bibliography{articles_trunc}

\end{document}